\documentclass[prd,twocolumn,showpacs,floatfix,amsmath,amssymb,floatfix]{revtex4}
\usepackage{graphicx,color,dcolumn,booktabs,bm,mathrsfs}
\usepackage{longtable,lscape}
\usepackage{txfonts}
\usepackage{overpic}
\usepackage{amssymb}
\usepackage{makecell}
\usepackage{indentfirst}
\usepackage{feynmf}   %{feynmp}
\usepackage{slashed}  %for Feynman symbols
\usepackage{cases}
\usepackage{color}
\usepackage{multirow}
\usepackage{epstopdf}
\usepackage{amssymb}% http://ctan.org/pkg/amssymb
\usepackage{pifont}% http://ctan.org/pkg/pifont
\newcommand{\cmark}{\ding{51}}%
\newcommand{\xmark}{\ding{55}}%

\graphicspath{{Figures/}} %
\usepackage[colorlinks,
                      citecolor=blue,
                      anchorcolor=red,
                      menucolor=red,
                      linkcolor=red,
                      filecolor=red,
                      runcolor=red,
                      urlcolor=blue,
                      frenchlinks=red]{hyperref}

\begin{document}

\title{A Phenomenological analysis on isospin-violating decay of $X(3872)$}
\author{Qi Wu$^{1}$}\email{wuq@seu.edu.cn}
\author{Dian-Yong Chen$^{1}$}\email{chendy@seu.edu.cn (corresponding author)}
\author{Takayuki Matsuki $^{2}$}\email{matsuki@tokyo-kasei.ac.jp}
\affiliation{$^1$School of Physics, Southeast University, Nanjing 210094, China\\
$^2$Tokyo Kasei University, 1-18-1 Kaga, Itabashi, Tokyo 173-8602, Japan
}

\begin{abstract}
In a molecular scenario, we investigate the isospin-breaking hidden charm decay processes of $X(3872)$, i.e., $X(3872) \to \pi^+ \pi^- J/\psi$, $X(3872) \to \pi^+ \pi^- \pi^0 J/\psi$, and $X(3872)\to  \pi^0\chi_{cJ}$. We assume that the source of the strong isospin violation comes from the different coupling strengths of $X(3872)$ to its charged components $D^{\ast +} D^-$ and neutral components $D^{\ast 0 } \bar{D}^0$ as well as the interference between the charged meson loops and neutral meson loops. The former effect could fix our parameters by using the measurement of the ratio $\Gamma[X(3872) \to \pi^+ \pi^- \pi^0 J/\psi]/\Gamma[X(3872) \to \pi^+ \pi^- J/\psi]$. With the determined parameter range, we find that the estimated ratio $\Gamma[X(3872) \to \pi^0 \chi_{c1}/\Gamma[X(3872) \to \pi^+ \pi^- J/\psi]$  is well consistent with the experimental measurement from the BESIII collaboration. Moreover, the partial width ratio of $\pi^0 \chi_{cJ}$ for $J=0,1,2$ is estimated to be $1.77\sim1.65:1:1.09\sim1.43$, which could be tested by further precise measurements of BESIII and Belle II.
\end{abstract}

\date{\today}
\maketitle

%%%%%%%%%%%%%%%%%%%%%%%%%%%%%%%%%%
\section{Introduction}
\label{sec:introduction}

As the first observed charmonium-like state, $X(3872)$ has been comprehensively investigated both from experimental and theoretical sides. It was first observed in $\pi^+ \pi^- J/\psi$ invariant mass spectrum of $B\to K \pi^+\pi^- J/\psi$ process by the Belle Collaboration in 2003~\cite{Choi:2003ue}. The mass and width were measured to be $m=3872.0 \pm 0.6 (\mathrm{stat.}) \pm 0.5 (\mathrm{syst.})\ \mathrm{MeV}$ and $\Gamma<2.3\ \mathrm{MeV}\ (90\% \ \mathrm{C.L.})$, respectively. The measured mass is very close to the thresholds of $D^\ast \bar{D}$, which are $3871.63$ MeV and $3879.91$ MeV for charged and neutral charmed meson pairs, respectively. Another interesting property of $X(3872)$ is its narrow width. These particular properties stimulated physicists with great interests in the nature of $X(3872)$.     After the observation by the Belle Collaboration, the charmonium-like state $X(3872)$ has been confirmed by the BaBar\cite{Aubert:2004fc,Aubert:2004ns,Aubert:2005eg,Aubert:2005zh,Aubert:2005vi,Aubert:2006aj,Aubert:2007rva,Aubert:2008gu,Aubert:2008ae,delAmoSanchez:2010jr}, CDF\cite{Acosta:2003zx,Abulencia:2005zc,Abulencia:2006ma,Aaltonen:2009vj}, D0\cite{Abazov:2004kp}, CMS\cite{CMS:2011yra,Vesentini:2012lea,Chatrchyan:2013cld,DallOsso:2013rtt,DallOsso:2014cmg,Sirunyan:2020qir}, LHCb\cite{Aaij:2011sn,LHCb:2011bia,LHCb:2011cra,Aaij:2013zoa,Aaij:2013rha,Aaij:2014ala,Aaij:2015eva,Aaij:2016kxn,Aaij:2017tzn,Aaij:2019zkm,Durham:2020zuw,Aaij:2020qga,
Aaij:2020xjx,Aaij:2020tzn}, and BESIII\cite{Ablikim:2013dyn,Ablikim:2019soz,Ablikim:2019zio,Ablikim:2020xpq} Collaborations in the invariant mass spectra of $\pi^+ \pi^- J/\psi$\cite{Choi:2003ue,Adachi:2008te,Aubert:2004ns,Aubert:2005eg,Aubert:2005zh,Aubert:2008gu,Acosta:2003zx,Abulencia:2005zc,Abulencia:2006ma,Aaltonen:2009vj,
Abazov:2004kp,CMS:2011yra,Vesentini:2012lea,Chatrchyan:2013cld,DallOsso:2013rtt,DallOsso:2014cmg,Sirunyan:2020qir,Aaij:2011sn,LHCb:2011bia,LHCb:2011cra,Aaij:2013zoa,
Aaij:2013rha,Aaij:2019zkm,Durham:2020zuw,Aaij:2020qga,Aaij:2020xjx,Aaij:2020tzn,Ablikim:2013dyn}, $\pi^+ \pi^- \pi^0 J/\psi$ \cite{Abe:2005ix,delAmoSanchez:2010jr,Ablikim:2019zio}, $D^{\ast 0} \bar{D}^0$\cite{Adachi:2008sua,Aubert:2007rva,Ablikim:2020xpq}, $D^{0} \bar{D}^0 \pi^0$\cite{Gokhroo:2006bt,Aubert:2005vi}, $\chi_{cJ} \pi^0$\cite{Ablikim:2019soz}, $\gamma J/\psi$\cite{Abe:2005ix,Aubert:2006aj,Aubert:2008ae,Ablikim:2020xpq}, and $\gamma \psi(2S)$\cite{Aubert:2008ae,Aaij:2014ala,Ablikim:2020xpq}. The quantum number of $X(3872)$ has been determined to be $I^G(J^{PC})=0^+ (1^{++})$ by the LHCb collaboration~\cite{Aaij:2013zoa}. In Table~\ref{Table:Exp1}, we have collected the experimental observations of $X(3872)$.

Beside the resonance parameters, the experimental measurements indicated that the $\pi^+ \pi^-$ invariant mass for $X(3872) \to \pi^+\pi^- J/\psi$ concentrates near the upper kinematic boundary, which corresponds to the $\rho$ meson mass~\cite{Choi:2003ue}. As for $X(3872) \to \pi^+ \pi^- \pi^0 J/\psi$, the $\pi^+ \pi^- \pi^0 $ invariant mass distribution has a strong peak between 750 MeV and the kinematic limit of 775 MeV, suggesting that the process is dominated by the sub-threshold decay $X(3872)\to \omega J/\psi$. The ratio of the branching fractions of $\pi^+ \pi^- J/\psi$ and $\pi^+ \pi^- \pi^0 J/\psi$ is determined to be~\cite{Abe:2005ix, Ablikim:2019zio,delAmoSanchez:2010jr},
\begin{eqnarray}
\renewcommand\arraystretch{1.35}
\frac{\mathcal{B}[X\to  J/\psi\pi^+ \pi^- \pi^0]}{\mathcal{B}[X\to  J/\psi\pi^+ \pi^-]}=
\left\{
\begin{array}{ll}
	1.0\pm 0.4 \pm 0.3  & \mathrm{Belle}\\
	1.43^{+0.28}_{-0.23} & \mathrm{BESIII} \\
	0.8\pm0.3 & \mathrm{BABAR}\\
\end{array}
\right.
\label{Eq:Ratio1}
\end{eqnarray}
The large isospin violation implied by the almost equality of the branching fractions of  $\omega J/\psi$ and $\rho J/\psi$ channels  further makes the nature of $X(3872)$ complicated and confusing. To understand the particular properties of $X(3872)$, a large number of attempts were made to reveal its nature. Since the mass of $X(3872)$ is very close to the threshold of $D\bar{D}^\ast$, it is natural to consider it as an analogue of the deuteron, i.e., $D\bar{D}^\ast$ loosely bound state, which was supported by the spectrum estimation in potential model \cite{Voloshin:1976ap,DeRujula:1976zlg,Tornqvist:1993ng,Tornqvist:2004qy,Wong:2003xk,Swanson:2003tb,Thomas:2008ja,Lee:2009hy,Chen:2009zzi,Gamermann:2009uq,Ortega:2010qq,
Li:2012cs,Guo:2013sya,Wang:2013kva}, by the QCD sum rule \cite{Wang:2013daa}, the decay and production property investigations \cite{Voloshin:2003nt,Voloshin:2004mh,Swanson:2004pp,Braaten:2003he,Close:2003sg,Braaten:2004rw,Braaten:2004ai,Braaten:2005jj,Braaten:2005ai,AlFiky:2005jd,Voloshin:2005rt,
Braaten:2006sy,Dubynskiy:2006cj,Fleming:2007rp,Braaten:2007ct,Liu:2006df,Braaten:2007sh,Fleming:2008yn,Dong:2008gb,Braaten:2007ft,Bignamini:2009sk,Canham:2009zq,
Harada:2010bs,Mehen:2011ds,Fleming:2011xa,Aceti:2012cb,Margaryan:2013tta,Guo:2013nza,Guo:2014hqa,Guo:2014yna,Guo:2014taa,Takeuchi:2014rsa,Voloshin:2019ivc,Zhou:2019swr,
Braaten:2019gfj}. However, the estimation in a chiral quark model disfavors the $S-$wave $D\bar{D}^\ast$ molecular interpretation even if all the possible meson exchanges were taken into account~\cite{Liu:2008qb}. In Refs.~\cite{Bugg:2004rk,Bugg:2004sh}, the authors indicated that the interaction of $D\bar{D}^\ast$ may not be strong enough to form a bound state, but their interaction can still lead to a cusp structure near the threshold.

\begin{table*}[!hbt]
\centering
\caption{\normalsize The experimental measurements of $X(3872)$ from different experiments, where \cmark denotes that the decay channel was observed but the mass of $X(3872)$ was not reported and \xmark\ indicates that channel was not observed.
 \label{Table:Exp1}}
 \renewcommand\arraystretch{1.35}
\begin{tabular}{p{2.5cm}<\centering |p{3.5cm}<\centering|p{3 cm}<\centering |p{3.5cm}<\centering|p{3cm}<\centering }
 \toprule[1pt]
 Experiments &Channel & Mass (MeV)&Channel & Mass (MeV) \\
 \midrule[1pt]
%\midrule[1pt]
\multirow{4}{*}{Belle}
& {$\pi^+ \pi^- J/\psi$\cite{Choi:2003ue}} & {$3872.0\pm0.78$}
& {$D^0 \bar{D}^0 \pi^0 $\cite{Gokhroo:2006bt}} & {$3875.2^{+1.1}_{-1.9}$}  \\
\cline{2-5}
& {$\pi^+ \pi^- J/\psi$\cite{Adachi:2008te}} & {$3871.46\pm0.38$}
& {$D^{\ast0}\bar{D}^0$\cite{Adachi:2008sua}} & {$3872.9^{+0.58}_{-0.78}$}  \\
\cline{2-5}
& \makecell*[c]{$ \gamma J/\psi$ \\ $\pi^+ \pi^- \pi^0 J/\psi$\cite{Abe:2005ix}} & {\cmark}&& \\
\hline
\multirow{8}{*}{BaBar}
& {$\eta J/\psi$~\cite{Aubert:2004fc}} & {\xmark}
& {$\pi^+ \pi^- J/\psi$\cite{Aubert:2004ns}} & {$3873.4\pm1.4$}   \\
\cline{2-5}
& {$\pi^+ \pi^- J/\psi$\cite{Aubert:2005eg}} & {\cmark}
& {$\pi^+ \pi^- J/\psi$\cite{Aubert:2005zh}} & \makecell*[c]{$3871.3\pm0.6(B^-)$\\$3868.6\pm1.2(B^0)$} \\
\cline{2-5}
& {$D^0 \bar{D}^0 \pi^0 $\cite{Aubert:2005vi}} & {\cmark}
& {$\gamma J/\psi$\cite{Aubert:2006aj}} & {\cmark}   \\
\cline{2-5}
& {$D^{\ast0}\bar{D}^0$\cite{Aubert:2007rva}} & {$3875.1^{+0.7}_{-0.9}$}
& {$\pi^+ \pi^- J/\psi$\cite{Aubert:2008gu}} & \makecell*[c]{$3871.4\pm0.6(B^+)$\\$3868.7\pm1.6(B^0)$}  \\
\cline{2-5}
& \makecell*[c]{$\gamma J/\psi$\\$\gamma J/\psi^\prime$\cite{Aubert:2008ae}} & {\cmark}
& {$\pi^+ \pi^- \pi^0 J/\psi$\cite{delAmoSanchez:2010jr}} & {$3873.0^{+2.2}_{-2.1}$}  \\
\cline{2-5}
\hline
\multirow{2}{*}{CDF}
& {$\pi^+ \pi^- J/\psi$\cite{Acosta:2003zx}} & {$3871.3\pm0.8$}
& {$\pi^+ \pi^- J/\psi$\cite{Abulencia:2005zc}} & {\cmark}   \\
\cline{2-5}
& {$\pi^+ \pi^- J/\psi$\cite{Abulencia:2006ma}} & {\cmark}
& {$\pi^+ \pi^- J/\psi$\cite{Aaltonen:2009vj}} & {$3871.61\pm0.25$} \\
\cline{2-5}
\hline
\multirow{1}{*}{D0}
& {$\pi^+ \pi^- J/\psi$\cite{Abazov:2004kp}} & {$3871.8\pm4.3$}  \\
\cline{2-3}
\hline
\multirow{3}{*}{CMS}
& {$\pi^+ \pi^- J/\psi$\cite{CMS:2011yra}} & {$3870.2\pm1.9$}
& {$\pi^+ \pi^- J/\psi$\cite{Vesentini:2012lea}} & {\cmark} \\
\cline{2-5}
& {$\pi^+ \pi^- J/\psi$\cite{Chatrchyan:2013cld}} & {\cmark}
& {$\pi^+ \pi^- J/\psi$\cite{DallOsso:2013rtt}} & {\cmark}  \\
\cline{2-5}
& {$\pi^+ \pi^- J/\psi$\cite{DallOsso:2014cmg}} & {\cmark}
& {$\pi^+ \pi^- J/\psi$\cite{Sirunyan:2020qir}} & {\cmark} \\
\cline{2-5}
\hline
\multirow{7}{*}{LHCb}
& {$\pi^+ \pi^- J/\psi$\cite{Aaij:2011sn}} & {$3871.95\pm0.49$}
& {$\pi^+ \pi^- J/\psi$\cite{LHCb:2011bia}} & {\cmark}  \\
\cline{2-5}
& {$\pi^+ \pi^- J/\psi$\cite{LHCb:2011cra}} & {$3871.96\pm0.47$}
& {$\pi^+ \pi^- J/\psi$\cite{Aaij:2013zoa}} & {\cmark} \\
\cline{2-5}
& {$\pi^+ \pi^- J/\psi$\cite{Aaij:2013rha}} & {\cmark}
& {$\gamma \psi^\prime$\cite{Aaij:2014ala}} & \makecell*[c]{$ 3873.4\pm3.4(J/\psi)$\\$3869.5\pm3.4(\psi^\prime)$} \\
\cline{2-5}
& {$\rho^0 J/\psi$\cite{Aaij:2015eva}} & {\cmark}
& {$p\bar{p}$\cite{Aaij:2016kxn}} & {\xmark}   \\
\cline{2-5}
& {$\phi\phi$\cite{Aaij:2017tzn}} & {\xmark}
& {$\pi^+ \pi^- J/\psi$\cite{Aaij:2019zkm}} & {\cmark}  \\
\cline{2-5}
& {$\pi^+ \pi^- J/\psi$\cite{Durham:2020zuw}} & {\cmark}
& {$\pi^+ \pi^- J/\psi$\cite{Aaij:2020qga}} & {$3871.695\pm0.095$}  \\
\cline{2-5}
& {$\pi^+ \pi^- J/\psi$\cite{Aaij:2020xjx}} & {$ 3871.59\pm0.07$}
& {$\pi^+ \pi^- J/\psi$\cite{Aaij:2020tzn}} & {\cmark}  \\
\cline{2-5}
\hline
\multirow{4}{*}{BESIII}
& {$\pi^+ \pi^- J/\psi$\cite{Ablikim:2013dyn}} & {$3871.9\pm0.73$}
& {$\pi^0 \chi_{c1}(1P)$\cite{Ablikim:2019soz}} & {\cmark} \\
\cline{2-5}
& {$\omega J/\psi$\cite{Ablikim:2019zio}} & {$3873.3\pm1.5$}
& \makecell*[c]{$D^{\ast0}\bar{D}^0+c.c.$\\$\gamma J/\psi$\\$\gamma \psi(2S)$\\$\gamma D^+ D^-$ \cite{Ablikim:2020xpq}} & \makecell*[c]{\cmark \\ \cmark \\ \xmark \\ \xmark} \\
\cline{2-5}
\hline
\bottomrule[1pt]
\end{tabular}
\end{table*}

It should be noticed that in all the observed decay modes of $X(3872)$, the final states contain charm and anti-charm quarks. Thus, the quark components of $X(3872)$ include at least $c\bar{c}$, which implies that $X(3872)$ could be a good charmonium candidate. Considering the $J^{PC}$ quantum number and the mass of $X(3872)$, one can assign it only as $\chi_{c1}(2P)$ state~\cite{Barnes:2003vb,Eichten:2004uh,Chen:2007vu,Meng:2007cx,Liu:2007uj,Wang:2010ej,Kalashnikova:2010hv,Wang:2012cp}. However, the resonance parameters and decay behaviors make us hardly interpret it in the charmonium scenario. Then, the tetraquark interpretation with constituent $c\bar{c}q\bar{q}$~\cite{Vijande:2004vt,Maiani:2005pe,Ebert:2005nc,Navarra:2006nd,Cui:2006mp,Matheus:2006xi,Nielsen:2006jn,Dubnicka:2010kz,Dubnicka:2011mm, Maiani:2004vq,Wang:2013vex} and hybrid with constituent $c\bar{c}g$\cite{Close:2003mb,Li:2004sta,Petrov:2005tp} have been proposed.

To date, the nature of $X(3872)$ still remains unclear. Besides the mass spectrum, the investigations of decay behaviors are also crucial to understand the properties of $X(3872)$. The isospin breaking effects of $X(3872)$ were studied in Ref.~\cite{Gamermann:2009fv}, where $X(3872)$ was considered as a dynamically generated state and the coupling strengths of $X(3872)D^{\ast +} D^{-}$ and $X(3872)D^{\ast 0} \bar{D}^{0}$ were assumed to be the same. Under $2^3P_1$ assignment, the decay channel $X(3872)\to  \rho/\omega +J/\psi$ was estimated via intermediate charmed meson loops~\cite{Meng:2007cx}. Using a phenomenological Lagrangian approach, the authors studied radiative decays to $J/\psi/\psi(2S)$ with the $X(3872)$ being a composite state containing both $D^0 \bar{D}^{\ast0}$ molecule and a $c\bar{c}$ component~\cite{Dong:2008gb,Dong:2009uf}, and hidden charm and radiative decays of $X(3872)$ were investigated with the $X(3872)$ being a composite state comprised of the dominant molecular $D^0 \bar{D}^{\ast0}$ component and other hadronic pairs, which could be $D^\pm D^{\ast\mp}$ and $J/\psi \omega/\rho$~\cite{Dong:2009yp}. The final state interaction effects of the hidden charm decay of $X(3872)$ were examined in Ref.~\cite{Liu:2006df}, they found that the FSI contribution to $X(3872)\rightarrow J/\psi \rho$ is tiny. Assuming the decays of $X(3872)$ through $\rho J/\psi$ and $\omega J/\psi$, the authors in Ref.~\cite{Braaten:2005ai} calculated the decay rates of $X(3872)\to  \pi^+ \pi^- J/\psi$ and $X(3872)\to  \pi^+ \pi^- \pi^0 J/\psi$.

Besides $\pi^+ \pi^- J/\psi$ and $\pi^+ \pi^- \pi^0 J/\psi$, the pionic transition from $X(3872)$ to $\chi_{cJ}$ was predicted in Refs.~\cite{Dubynskiy:2007tj,Fleming:2008yn}, in which it is found that the ratio of different transitions with different angular momentum $J$ was sensitive to the inner structure of $X(3872)$. In 2019, the BESIII Collaboration searched for the process $e^+ e^- \to \gamma X(3872)$ by using the collision data with center-of-mass energies between 4.15 and 4.30 GeV and a new decay mode, $\chi_{c1} \pi^0$, of $X(3872)$ was observed with a statistical significance of more than 5$\sigma$ but no significant $X(3872)$ signal was observed in the invariant mass distributions of $\pi^0 \chi_{c0,2}$. The ratios of the branching ratios of $X(3872)\to \pi^0\chi_{cJ} $ for $J=0,1,2$ and $X(3872) \to \pi^+ \pi^- J/\psi$ were measured to be~\cite{Ablikim:2019soz},
\begin{eqnarray}
\renewcommand\arraystretch{1.35}
\frac{\mathcal{B}[X \to \pi^0 \chi_{cJ} ]}{\mathcal{B}[X\to \pi^+ \pi^- J/\psi]} = \left\{
\begin{array}{ll}
6.6^{+6.5}_{-4.5} \pm 1.1~(19) & J=0\\
0.88^{+0.33}_{-0.27} \pm 0.10 & J=1\\
0.40^{+0.37}_{-0.27} \pm 0.04~(1.1) & J=2\\	
\end{array}
\right.
\label{Eq:Ratio2}
\end{eqnarray}
where the numbers in parentheses for $J=0$ and $2$ are the upper limits in $90\%$ C.L.~\cite{Ablikim:2019soz}. Later, the Belle collaboration searched for $X(3872)$ in $B^+ \to \chi_{c1} \pi^0 K^+$ decay, and the ratio was measured to be $\mathcal{B}[X \to \pi^0 \chi_{c1} ]/\mathcal{B}[X\to \pi^+ \pi^- J/\psi]<0.97$ at $90\%$ C.L.~\cite{Bhardwaj:2019spn}.  The upper limit of the ratio measured by the Belle collaboration does not contradict the BESIII data~\cite{Ablikim:2019soz}.

The experimental measurements of the ratios in Eqs.~(\ref{Eq:Ratio1}) and (\ref{Eq:Ratio2}) imply a strong isospin violation. The explanation of this fact is important for revealing the nature of $X(3872)$. In the present work, we attempt to hunt for the source of the isospin violation in the molecular scenario by assuming that $X(3872)$ is an $S$-wave molecule with $J^{PC}=1^{++}$ given by the superposition of $D^0{\bar D}^{*0}$ and $D^\pm D^{\ast\mp}$ hadronic configurations. The fundamental source of the isospin violation is the mass difference of up and down quarks. Specific to the present discussed issue, the concrete manifestation is the mass difference of charged and neutral charmed mesons, which leads to the different coupling strengths of  $X(3872) D^{\ast 0} \bar{D}^0$ and $X(3872) D^{\ast +} D^-$. This coupling strength difference in part provides the source of the isospin violation in the decays of $X(3872)$. As a molecular state, the hidden charm decays of $X(3872)$ occur via the charmed meson loops, where the interferences between the charged and neutral  meson loops provide another important source of the isospin violation. In the present work, we consider these two sources of isospin violation, the uncertainties of the former one, i.e. the different coupling strengths, can be determined by the ratio $\mathcal{B}[X\to \pi^+ \pi^- \pi^0 J/\psi]/\mathcal{B}[X\to \pi^+ \pi^- J/\psi]$, and then with the fixed parameters, we can further estimate the ratios $\mathcal{B}[X\to \pi^0 \chi_{cJ}]/\mathcal{B}[X\to \pi^+ \pi^- J/\psi]$. Comparing the present estimation for $J=1$ with the BESIII data can also check the present model's reasonability. As for $J=0$ and $J=2$, the present estimations can narrow down the ratios' range, which could be tested by further measurements.

This paper organized as follows: After introduction, we present the model used in the present estimations of $X(3872)\to  \rho/\omega J/\psi$ and $X(3872)\to  \chi_{cJ}\pi^0$. The numerical results and discussions are presented in  Sec.~\ref{sec:results}, and Sec.~\ref{sec:summary} is devoted to a short summary.

\section{hidden charm decay of $X(3872)$}
\label{sec:decay}

\begin{figure}[htb]
\begin{tabular}{ccc}
  \centering
 \includegraphics[width=4.2cm]{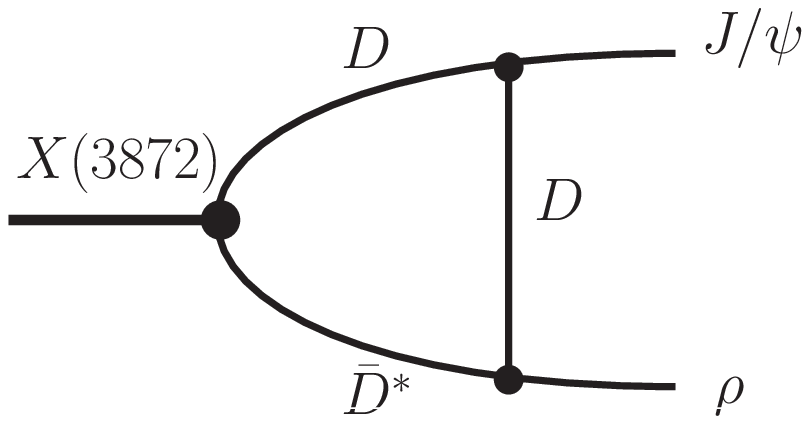}&
 \includegraphics[width=4.2cm]{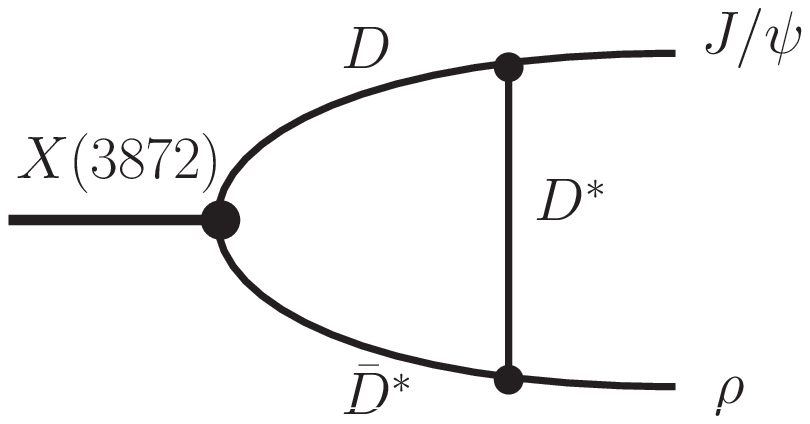}\\
 \\
 $(a)$ & $(b)$ \\
 \\
 \includegraphics[width=4.2cm]{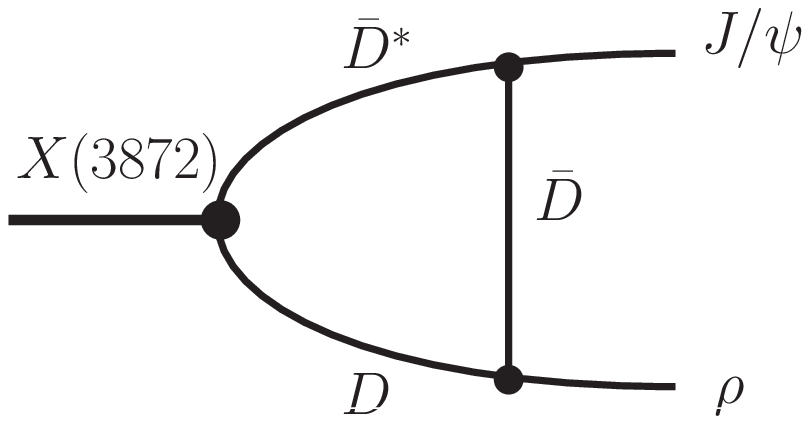}&
 \includegraphics[width=4.2cm]{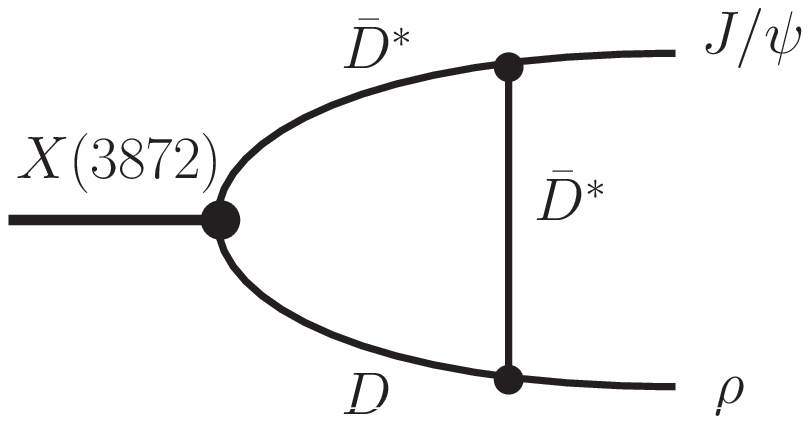}&\\
 \\
 $(c)$ & $(d)$&\\
 \end{tabular}
  \caption{Diagrams  contributing to $X(3872)\to  J/\psi \rho$.   The charge conjugate diagrams are not shown but included in the calculations.}\label{Fig:Tri1}
\end{figure}

\begin{figure}[htb]
\begin{tabular}{ccc}
  \centering
 \includegraphics[width=4.2cm]{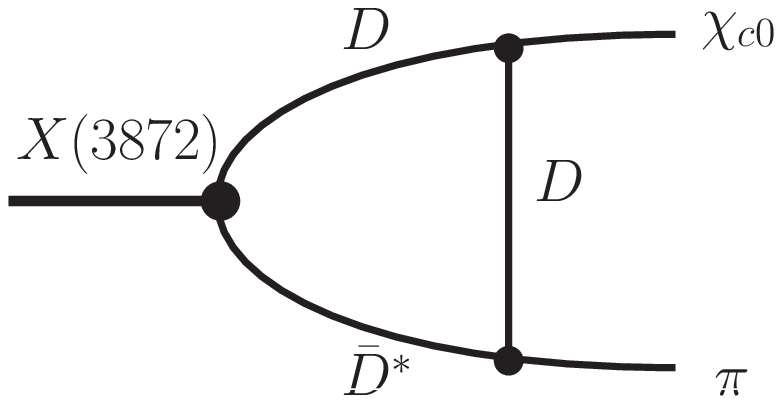}&
 \includegraphics[width=4.2cm]{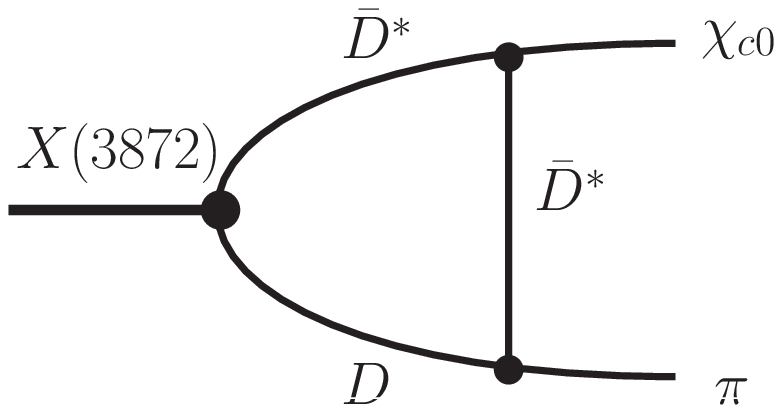}\\
 \\
 $(a)$ & $(b)$ \\
 \\
 \includegraphics[width=4.2cm]{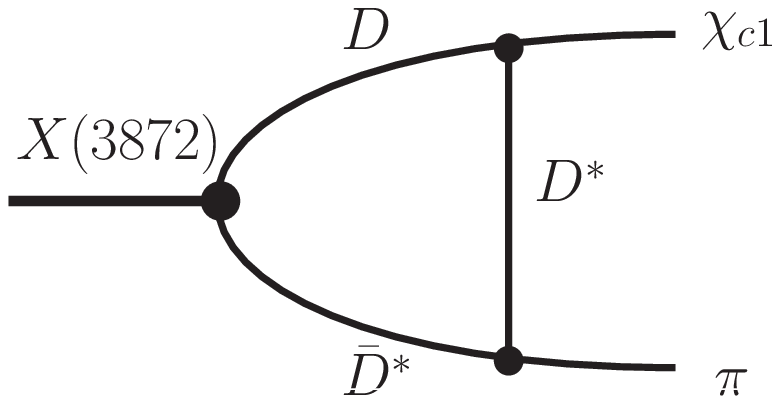}&
 \includegraphics[width=4.2cm]{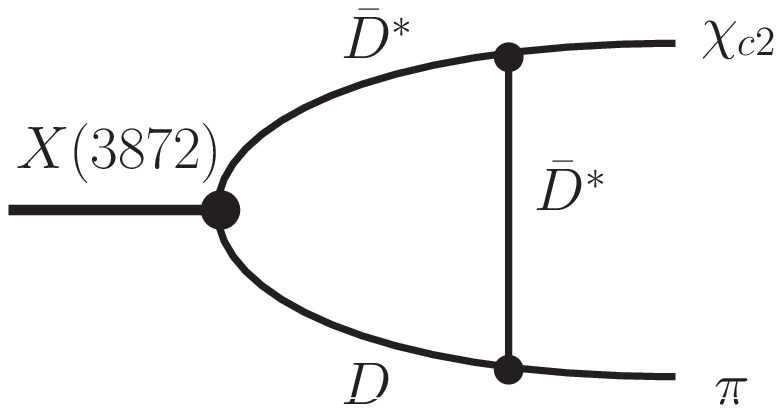}&\\
 \\
 $(c)$ & $(d)$&\\
 \end{tabular}
  \caption{Diagrams  contributing to $X(3872)\to  \chi_{cJ}\pi$, the charge conjugate diagrams are not shown but included in the calculations.}\label{Fig:Tri2}
\end{figure}

As discussed above, we assume that the coupling strengths of $X(3872) D^{\ast 0} \bar{D}^0$ and $X(3872) D^{\ast +} D^-$ are different. The effective coupling of $X(3872)$ with its components can be,
\begin{eqnarray}
{\cal L}_{X(3872)}&=&\frac {g_X} {\sqrt {2}}X^{\dagger}_\mu \Big[\sin\theta\left( D^{*0\mu} {\bar D}^0 + D^0 {\bar D}^{*0\mu} \right )\nonumber \\
&&+\cos\theta \Big(D^{*+\mu} D^- + D^+ D^{*-\mu}\Big)\Big]\label{eq:lagX}
\end{eqnarray}
where $g_X$ is the coupling constant, $\theta$ is a phase angle describing the proportion of neutral and charged constituents.

It should be mention that from a more fundamental quark level point of view, the difference in coupling strength should be dynamically generated from $u,\ d$ quark-antiquark pair generating processes and also from the wave function difference of the charged and neutral $D\bar{D}^\ast$. Furthermore, from the dispersion discussion in Refs.~\cite{Suzuki:2005ha, Meng:2007cx}, the different coupling strength comes also from the dispersion integrals where the thresholds are different for the charged and neutral channels. From the phenomenological point of view, we can parameterize the coupling strength and the isospin breaking effects into a common factor $g_X$  and a phase angle $\theta$ in the effective Lagrangian as shown in Eq.~(\ref{eq:lagX}).   

Moreover, the distributions of the components, i.e., $D\bar{D}^\ast$, in the molecular state could be described by a wave function, which would then be integrated in the Feynman diagram calculations and affect the magnitude of partial widths. In the present work, we mainly focus on the ratios of the partial widths as given in Eqs. (\ref{Eq:Ratio1})-(\ref{Eq:Ratio2}), the form factor appears in both the numerators and denominators. As discussed in Ref.~\cite{Dong:2008gb}, the estimated ratio in the nonlocal case may not be too much different from the local one. Then the simple parameterization in Eq.~(\ref{eq:lagX}) could be a reasonable approximation in estimating the order of magnitude of the ratio.

In the present work, the hidden charm decay processes of $X(3872)$ occur via charmed meson loops, i.e., the charmonium and light meson in the final state couple to the components of $X(3872)$ by exchanging a proper charmed meson. The diagrams contributing to $X(3872) \to \rho J/\psi$ are presented in Fig.~\ref{Fig:Tri1}. The diagrams contributing to $X(3872) \to \omega J/\psi$ could be obtained by replacing the $\rho$ meson by the $\omega$ meson, while the diagrams contributing to $X(3872) \to \pi^0 \chi_{cJ}$ with $J=0,1,2$ are presented in Fig.~\ref{Fig:Tri2}. Here, we do not include the the four-meson interactions, such as $D^\ast D J/\psi \rho$, which is equivalent to the $D D^\ast $ coupling to $J/\psi \rho$ through an exchange of an excited charmed meson. In the intermediate meson loop model, the dominant contributions are usually supposed to come from the ground states of mesons which have strong coupling with the final states, while the contribution from the excited mesons is suppressed.

In the present work, all these diagrams are estimated in hadronic level and all the involved interactions are depicted by effective Lagrangians. In heavy quark limit, one can construct the effective Lagrangian for charmonium and charmed mesons, which are~\cite{Oh:2000qr,Colangelo:2002mj,Casalbuoni:1996pg}
\begin{eqnarray}
\mathcal{L}_{\psi D^{(\ast)} D^{(\ast)}}&=&-ig_{\psi DD}\psi_\mu(\partial^\mu DD^\dag-D\partial^\mu D^\dag)\nonumber\\
&&+g_{\psi D^{\ast}D}\varepsilon^{\mu\nu\alpha\beta}\partial_\mu \psi_\nu(D^\ast_\alpha \stackrel{\leftrightarrow}{\partial}_\beta D^\dag-D \stackrel{\leftrightarrow}{\partial}_\beta D_\alpha^{\ast\dag})\nonumber\\
&&+ig_{\psi D^{\ast}D^{\ast}}\psi^\mu(D^\ast_\nu \stackrel{\leftrightarrow}{\partial}^\nu D^{\ast\dag}_\mu\nonumber\\
&&+D^\ast_\mu \stackrel{\leftrightarrow}{\partial}^\nu D^{\ast\dag}_\nu-D^\ast_\nu \stackrel{\leftrightarrow}{\partial}_\mu D^{\ast\nu\dag}),\nonumber\\
\mathcal{L}_{\chi_{cJ} D^{(\ast)} D^{(\ast)}}&=&g_{\chi_{c0} D D}\chi_{c0} D_i D_i^\dag + g_{\chi_{c0} D^* D^*}\chi_{c0} D^{*}_{i\mu} D_i^{*\mu\dag}\nonumber\\
&&+ig_{\chi_{c1} D^{*} D} \chi_{c1}^\mu (D^*_{i\mu} D^{\dag}_i-D_i D^{\ast\dag}_{i\mu} ) \nonumber \\
&& +g_{\chi_{c2} D^{*} D^{*}} \chi_{c2}^{\mu \nu} D_{i\mu}^{*} D_{i\nu}^{* \dagger} \, ,\label{eq:charmonium}
\end{eqnarray}

The coupling between light meson and charmed mesons could be obtained  based on the heavy quark limit and chiral symmetry, which are ~\cite{Casalbuoni:1996pg,Colangelo:2003sa,Cheng:2004ru}
\begin{eqnarray}
 {\cal L} &=& -ig_{D^{\ast }D
{\mathcal P}}\left( D^i \partial^\mu {\mathcal P}_{ij} D_\mu^{\ast
j\dagger }-D_\mu^{\ast i}\partial^\mu {\mathcal P}_{ij} D^{j \dag}\right) \nonumber \\
&& +\frac{1}{2}g_{D^\ast D^\ast {\mathcal P}}\varepsilon _{\mu
\nu \alpha \beta }D_i^{\ast \mu }\partial^\nu {\mathcal P}^{ij}  {\overset{
\leftrightarrow }{\partial }}{\!^{\alpha }} D_j^{\ast \beta\dag } - ig_{{D}{D}\mathcal{V}} {D}_i^\dagger {\stackrel{\leftrightarrow}{\partial}}{\!_\mu} {D}^j(\mathcal{V}^\mu)^i_j \nonumber \\
&& -2f_{{D}^*{D}\mathcal{V}} \epsilon_{\mu\nu\alpha\beta}
(\partial^\mu \mathcal{V}^\nu)^i_j
({D}_i^\dagger{\stackrel{\leftrightarrow}{\partial}}{\!^\alpha} {D}^{*\beta j}-{D}_i^{*\beta\dagger}{\stackrel{\leftrightarrow}{\partial}}{\!^\alpha} {D}^j) \nonumber
\\
&&+ ig_{{D}^*{D}^*\mathcal{V}} {D}^{*\nu\dagger}_i {\stackrel{\leftrightarrow}{\partial}}{\!_\mu} {D}^{*j}_\nu(\mathcal{V}^\mu)^i_j \nonumber \\
&& +4if_{{D}^*{D}^*\mathcal{V}} {D}^{*\dagger}_{i\mu}(\partial^\mu \mathcal{V}^\nu-\partial^\nu
\mathcal{V}^\mu)^i_j {D}^{*j}_\nu +{\rm H.c.} , \label{eq:light-meson}
 \label{eq:LDDV}
 \end{eqnarray}
where the ${D}^{(\ast)\dagger}=(\bar{D}^{(\ast)0},D^{(\ast)-},D^{(\ast)-}_s)$ is the charmed meson triplet, $\mathcal P$ and ${\mathcal V}_\mu$ are $3\times 3$ matrices for the nonet pseudoscalar and nonet vector mesons, respectively,
\begin{eqnarray}
  \mathcal{P} &=&
 \left(
 \begin{array}{ccc}
\frac{\pi^0}{\sqrt{2}} + \alpha \eta + \beta \eta^\prime & \pi^{+} & K^{+}\\
\pi^{-} & -\frac{\pi^0}{\sqrt{2}}+ \alpha \eta + \beta \eta^\prime   &  K^{0}\\
 K^{-} & \bar{K}^{0} & \gamma \eta + \delta \eta^\prime
 \end{array}
 \right),\nonumber \\
%\end{eqnarray}
%\begin{eqnarray}
\mathcal{V} &=& \left(\begin{array}{ccc}\frac{\rho^0} {\sqrt {2}}+\frac {\omega} {\sqrt {2}}&\rho^+ & K^{*+} \\
\rho^- & -\frac {\rho^0} {\sqrt {2}} + \frac {\omega} {\sqrt {2}} & K^{*0} \\
K^{*-}& {\bar K}^{*0} & \phi \\
\end{array}\right) \, .
\end{eqnarray}

With the Lagrangian listed above, we can obtain the decay amplitude corresponding to $X(3872) \to \rho J/\psi, \omega J/\psi$, and $ \pi^0 \chi_{cJ}$ with $J=0,1,2$. For brevity, we collect all the amplitudes corresponding to diagrams in Figs.~\ref{Fig:Tri1} and \ref{Fig:Tri2} in Appendix.~\ref{sec:appendix} and leave the coupling constants to be discussed in the following section.

In the present estimation, since the threshold of  $J/\psi \rho$ is very close to the mass of $X(3872)$, the width of $\rho$ meson should be included, and then, the width of $X(3872) \to \rho J/\psi$ should be,
\begin{eqnarray}
\Gamma_{X(3872)\to  J/\psi \rho}&=&\frac{1}{W_\rho}\int^{(m_X-m_{J/\psi})^2}_{(2m_\pi)^2} ds f(s,m_\rho,\Gamma_\rho)\nonumber\\
&&\times\frac{|\vec{p}|}{24\pi m^2_X}\mid\overline{\mathcal{M}^{\mathrm{tot}}_{X(3872)\to  J/\psi \rho}(m_\rho\to \sqrt{s})}\mid^2 \nonumber\\ \label{Eq:Gamma}
\end{eqnarray}
where $W_\rho=\int^{(m_X-m_{J/\psi})^2}_{(2m_\pi)^2} ds f(s,m_\rho,\Gamma_\rho)$, $f(s,m_\rho,\Gamma_\rho)$ is a relativistic form of the Breit-Wigner distribution, which reads
\begin{eqnarray}
f(s,m_\rho,\Gamma_\rho)=\frac{1}{\pi}\frac{m_\rho \Gamma_\rho}{(s-m^2_\rho)^2+m^2_\rho \Gamma^2_\rho},
\end{eqnarray}
and the amplitude $\mathcal{M}^{\mathrm{tot}}_{X(3872)\to  J/\psi \rho}(m_\rho\to \sqrt{s})$ can be obtained by replacing the mass of $\rho$ meson by $\sqrt{s}$ in the amplitudes listed in the appendix, in the same way the momentum of the final state becomes,
\begin{eqnarray}
|\vec{p}|=\frac{\sqrt{[m^2_X-(\sqrt{s}-m_{J/\psi})^2][m^2_X-(\sqrt{s}+m_{J/\psi})^2]}}{2m_X}
\end{eqnarray}

As for $X(3872)\to  J/\psi \pi^+ \pi^- \pi^0$, it is a sub-threshold decay process, hence, the Breit-Wigner distributions of $\omega$ meson should be also considered in a similar way. However, the lower limit of integral in Eq.~(\ref{Eq:Gamma}) and $W_\rho$ should be replaced with $(3m_{\pi})^2$. With the partial widths of $X(3872) \to \omega /\rho  J/\psi$, one can obtain the partial widths of $X(3872) \to \pi^ +\pi^- J/\psi$ and $X(3872) \to \pi^ +\pi^-  \pi^0 J/\psi$, which are,
\begin{eqnarray}
&&\Gamma[X(3872)\to  J/\psi \pi^+ \pi^-]= \nonumber\\
&&\hspace{1cm}\Gamma[X(3872)\to  \rho^0 J/\psi ] \mathcal{B}[\rho^0  \to \pi^+ \pi^- ]\nonumber\\
&&\Gamma[X(3872)\to  J/\psi \pi^+ \pi^- \pi^0]=\nonumber\\
&&\hspace{1cm}\Gamma[X(3872)\to   \omega J/\psi] \mathcal{B}[\omega \to \pi^ +\pi^- \pi^0] \nonumber
\end{eqnarray}
where $\mathcal{B}[\rho^0 \to \pi^+ \pi^-]\simeq 100\%$ and $\mathcal{B}[\omega \to \pi^+ \pi^- \pi^0] = (89.3\pm 0.6) \% $ are the branching ratios of $\rho^0 \to \pi^+ \pi^-$ and $\omega \to \pi^+ \pi^- \pi^0$, respectively.

\section{Numerical results and discussion}
\label{sec:results}

Since mass difference of $X(3872)$ and $D^{\ast 0}\bar{D}^0$ is very tiny, the coupling constants $g_X$ are very sensitive to the mass of $X(3872)$. Thus, in the present work, we mainly focus on the ratios of the hidden charm decay channels, which are independent on the coupling constants $g_X$. Moreover, the involved charmonia in the present estimation are $J/\psi$ and $\chi_{cJ}$. In the heavy quark limit, the coupling constants of the involved charmonia and charmed mesons can be related to the gauge couplings $g_1$ and $g_2$ by,
\begin{eqnarray}
g_{\psi DD}&=&2g_1 \sqrt{m_{\psi}}m_D,\nonumber \\ g_{\psi D^\ast D} &=&2g_1 \sqrt{m_{\psi} m_{D^\ast}/m_D},\nonumber \\
g_{\psi D^\ast D^\ast}&=&2g_1 \sqrt{m_{\psi}}m_{D^\ast},\nonumber \\
g_{\chi_{c0}DD}&=&-2\sqrt{3}g_2 \sqrt{m_{\chi_{c0}}} m_D ,\nonumber \\
g_{\chi_{c0}D^\ast D^\ast}&=&-\frac{2}{\sqrt{3}}g_2 \sqrt{m_{\chi_{c0}}} m_{D^\ast}\nonumber \\
g_{\chi_{c1}D^*D}&=& 2 \sqrt{2} g_2 \sqrt{2m_{\chi_{c1}} m_D m_{D^*}}\nonumber \\ g_{\chi_{c2}D^* D^*} &=& 4g_2 \sqrt{m_{\chi_{c2}}} m_{D^*}\nonumber
\end{eqnarray}
where $g_1=\sqrt{m_\psi}/(2m_D f_\psi)$, $g_2=-\sqrt{{m_{\chi_{c0}}}/{3}}/{f_{\chi_{c0}}}$, and $f_\psi=426 {\rm MeV}$ and $f_{\chi_{c0}}=510 {\rm MeV}$ are the $J/\psi$ and $\chi_{c0}$ decay constants~\cite{Colangelo:2002mj}, respectively.

In the heavy quark and chiral limits, the charmed meson couplings to the light vector and pesudoscalar mesons have the following relationship~\cite{Casalbuoni:1996pg, Cheng:2004ru},
\begin{eqnarray}
g_{{ D}{ D}V} = g_{{ D}^*{ D}^*V}=\frac{\beta g_V}{\sqrt{2}} , \quad f_{{ D}^*{ D}V}=\frac{ f_{{ D}^*{ D}^*V}}{m_{{ D}^*}}=\frac{\lambda g_V}{\sqrt{2}} \, , \\
g_{{D}^{*} {D} \mathcal{P}}=\frac{2 g}{f_{\pi}} \sqrt{m_{{D}} m_{{D}^{*}}}, \quad g_{{D}^{*} {D}^{*} {P}}=\frac{g_{{ D}^{*} { D} {\mathcal {P}}}}{\sqrt{m_{{D}} m_{{D}^{*}}}}
\end{eqnarray}
where the parameter $g_V = {m_\rho /f_\pi}$ with $f_\pi = 132$ MeV being the pion decay constant and $\beta=0.9$ ~\cite{Casalbuoni:1996pg}. By matching the form factor obtains from the light cone sum rule with that calculated from lattice QCD, one obtained the parameters $\lambda = 0.56 \, {\rm GeV}^{-1} $ and $g=0.59$~\cite{Isola:2003fh}.

In the amplitudes, the form factors should be considered to depict the inner structures and off shell effects of the charmed mesons in the loop. However, the mass of $X(3872)$ is very close to the thresholds of $D^\ast D$, which indicates that the components of $X(3872)$, i.e., the charmed mesons connected to $X(3872)$ in Figs. \ref{Fig:Tri1}-\ref{Fig:Tri2}, are almost on shell. Therefore, we introduce only one form factor in a monopole form to depict the inner structure and the off-shell effects of the exchanged charmed meson, which is \cite{Cheng:2004ru, Tornqvist:1993vu, Tornqvist:1993ng, Locher:1993cc, Li:1996yn},
\begin{eqnarray}
\mathcal{F}\left(q^{2}\right)=\frac{m^{2}-\Lambda^{2}}{q^{2}-\Lambda^{2}}\label{Eq:FFs1}
\end{eqnarray}
where the parameter $\Lambda$ can be further reparameterized as $\Lambda_{D^{(\ast)}}=m_{D^{(\ast)}}+\alpha\Lambda_{\rm QCD} $ with $\Lambda_{\rm QCD}=0.22 \ {\rm GeV}$ and $m_{D^{(\ast)}}$ is the mass of the exchanged meson. The model parameter $\alpha$ should be of order of unity~\cite{Tornqvist:1993vu,Tornqvist:1993ng,Locher:1993cc,Li:1996yn}, but its concrete value cannot be estimated by the first principle. In practice, the value of $\alpha$ is usually determined by comparing theoretical estimates with the corresponding experimental measurements.

\begin{figure}[htb]
  \centering
 \includegraphics[width=9cm]{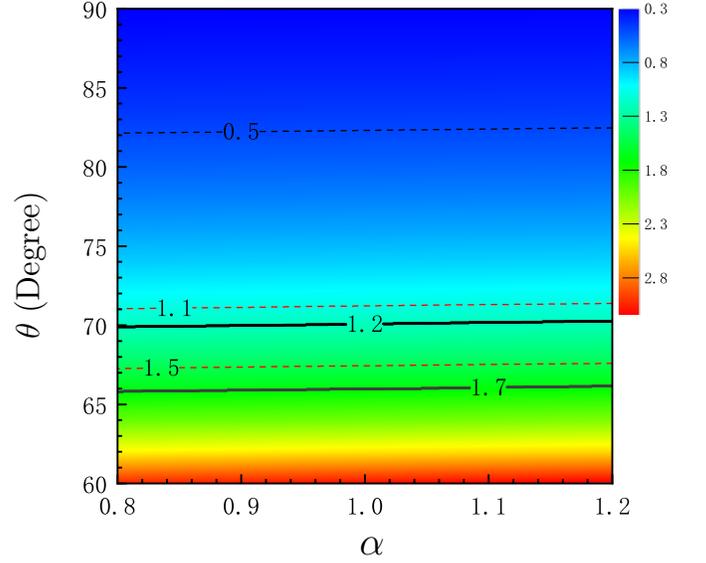}
  \caption{The ratio $\Gamma[X\to  J/\psi\pi^+ \pi^- \pi^0]/\Gamma[X\to  J/\psi\pi^+ \pi^- ]$ depending on the parameter $\theta$ and $\alpha$. The black solid lines are the upper and lower limit from the BESIII measurements, while the dashed lines are the upper and lower limits determined by BABAR and Belle collaborations.}\label{Fig:ratio1}
\end{figure}

In Fig.~\ref{Fig:ratio1}, we present the ratio  $\Gamma[X\to  J/\psi\pi^+ \pi^- \pi^0]/\Gamma[X\to  J/\psi\pi^+ \pi^- ]$ depending on the parameter $\theta$ and $\alpha$. For comparison, we also present the upper and lower limits measured by the BESIII, BABAR, and Belle collaborations~\cite{delAmoSanchez:2010jr, Ablikim:2019zio, Abe:2005ix}. From the figure, one can see the ratio is almost independent on the parameter $\alpha$ due to the similarity of $\rho J/\psi $ and $\omega J/\psi$ decay modes. Taking the latest BESIII data~\cite{Ablikim:2019zio} as a scale, the determined $\theta$ range is $66^\circ \sim 70^\circ$, which indicates that in the $X(3872)$, the weight of $D^0 \bar{D}^{\ast0}$ component is $(83 \sim 88)\%$. Such a large weight is expected since the threshold of the neutral component is very close to the mass of $X(3872)$~\cite{Gamermann:2009fv}.

\begin{figure}[htb]
  \centering
 \includegraphics[width=9cm]{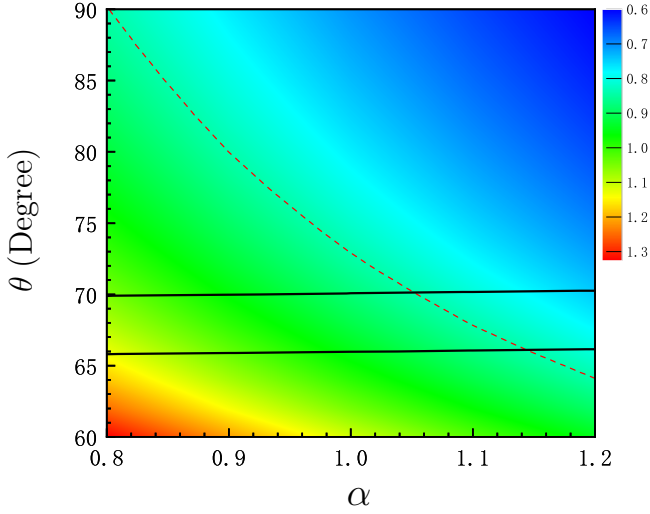}
  \caption{The ratio $\Gamma[X\to  \chi_{c1} \pi^0]/\Gamma[X\to  J/\psi \pi^+ \pi^-]$ depending on the parameter $\alpha$ and $\theta$. The red dashed curve indicates the center values of the ratio reported by the BESIII collaboration~\cite{Ablikim:2019soz}, while the black solid lines indicate the range determined by the ratio $\Gamma[X\to  J/\psi\pi^+ \pi^- \pi^0]/\Gamma[X\to  J/\psi\pi^+ \pi^- ]$.}\label{Fig:ratio2}
\end{figure}

\begin{figure}[htb]
  \centering
 \includegraphics[width=7.5cm]{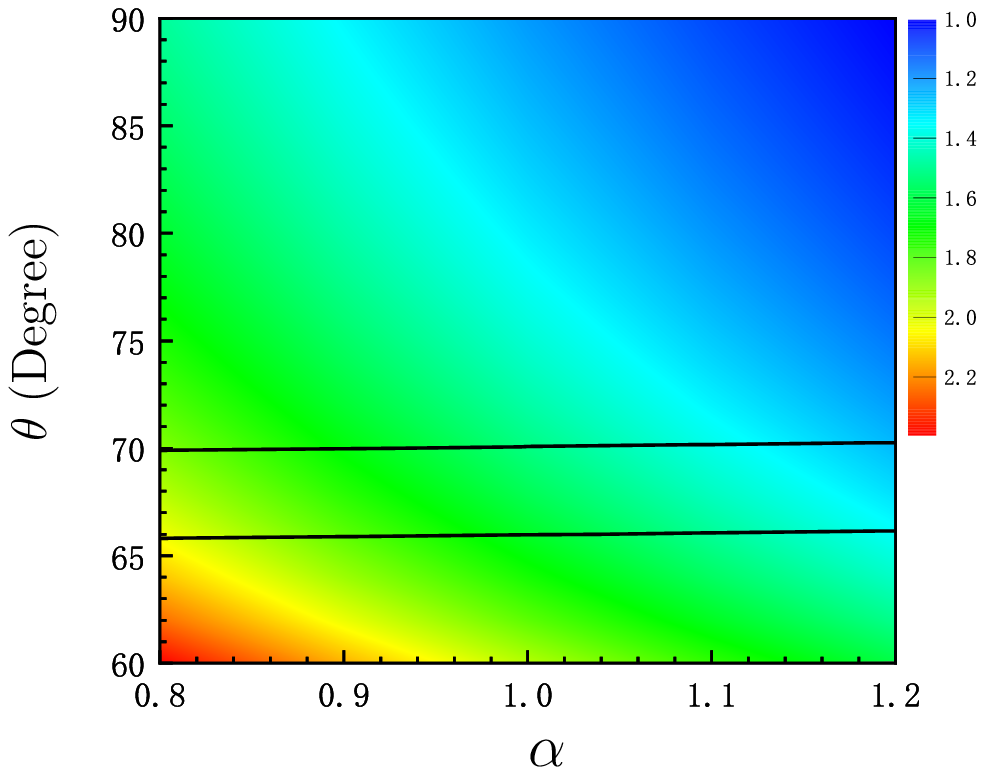}
  \caption{The same as Fig.~\ref{Fig:ratio2} but for $\Gamma[X\to \chi_{c0} \pi^0]/\Gamma[X\to J/\psi \pi^+ \pi^-]$.}\label{Fig:ratio3}
\end{figure}

\begin{figure}[htb]
  \centering
 \includegraphics[width=7.5	cm]{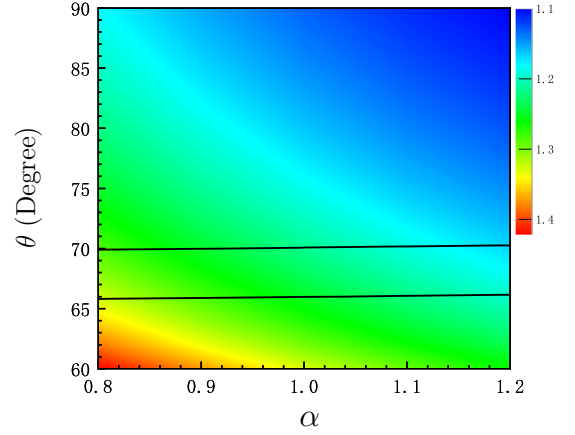}
  \caption{The same as Fig.~\ref{Fig:ratio2} but for $\Gamma[X\to \chi_{c2} \pi^0]/\Gamma[X\to J/\psi \pi^+ \pi^-]$.}\label{Fig:ratio4}
\end{figure}

As we have discussed in the introduction, there are two sources of the isospin violation in $X(3872)$ decays. When one takes $\theta=45^\circ$, the coupling strengths of $X(3872) D^{\ast 0} \bar{D}^{0}$ and $X(3872) D^{\ast +} D^-$ are the same. Hence, the isospin violation comes from the interference between the charged and neutral meson loops. In this case, the ratio $\Gamma[X \to \pi^+ \pi^- \pi^0 J/\psi]/\Gamma[X \to \pi^+ \pi^- J/\psi]$ is  estimated to be of the order of $10^{2}$, which indicates that the contributions from the interference between the charged and neutral meson loops are rather small and the dominant source of isospin violation in $X(3872)$ decays should come from the different coupling strengths of $X(3872) D^{\ast 0} \bar{D}^{0}$ and $X(3872) D^{\ast +} D^-$. In Refs.~\cite{Meng:2007cx, Suzuki:2005ha}, the authors assigned $X(3872)$ as a $2P$ charmoninum, and the large isospin violating came from the dispersion integrals where the mass of charged and neutral in the loops are different. In a $c\bar{c}$-two-meson hybrid model, the ratio $\Gamma[X(3872) \to \pi^+ \pi^- \pi^0 J/\psi] / \Gamma[X(3872) \to \pi^+ \pi^- J/\psi]$ was estimated to be $1.27 \sim 2.24$~\cite{Takeuchi:2014rsa}. In the $D^\ast \bar{D}$ molecular scenario, considering the $S-D$ mixing and the isospin mass splitting~\cite{Li:2012cs, Tornqvist:2004qy}, the ratio $\Gamma[X(3872) \to \pi^+ \pi^- \pi^0 J/\psi] / \Gamma[X(3872) \to \pi^+ \pi^- J/\psi]$ was evaluated as $0.42$~\cite{Li:2012cs}, which was close to the lower limit of the experimental data. In Ref.~\cite{Terasaki:2009in}, $X(3872)$ was considered as a tetraquark state and the isospin violating decay process $X(3872) \to \pi^+ \pi^- J/\psi$ was assumed to occur through $\rho^0 $ meson pole which is caused by the $\omega-\rho^0$ mixing.

In Fig.~\ref{Fig:ratio2}, we present the ratio $\Gamma[X\to  \chi_{c1} \pi^0]/\Gamma[X\to  J/\psi \pi^+ \pi^-]$ depending on the parameter $\alpha$ and $\theta$. In the $\theta$ range determined by $\Gamma[X\to \pi^+ \pi^- \pi^0 J/\psi]/\Gamma[X\to \pi^+ \pi^- J/\psi]$, we find the ratio $\Gamma[X\to  \chi_{c1} \pi^0]/\Gamma[X\to   \pi^+ \pi^-J/\psi]$ is determined to be $ 0.79 \sim 1.17 $, which is well consistent with the measurement of the BESIII collaboration, which is $\Gamma[X\to  \chi_{c1} \pi^0]/\Gamma[X\to  \pi^+ \pi^- J/\psi] =0.88^{+0.33}_{-0.27} \pm 0.10$~\cite{Ablikim:2019soz}.

\begin{figure}[t]
  \centering
 \includegraphics[width=8.5cm]{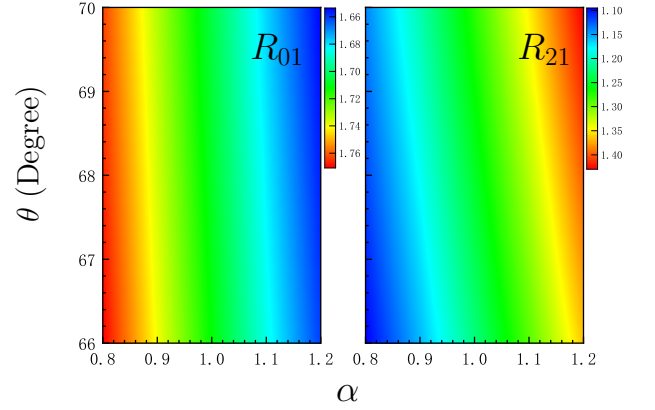}
  \caption{The ratios $R_{01}=\Gamma[X\to \pi^0 \chi_{c0}]/\Gamma[X\to \pi^0 \chi_{c1}] $ (left panel) and $R_{21}=\Gamma[X\to \pi^0 \chi_{c2}]/\Gamma[X\to \pi^0 \chi_{c1}] $ (right pannel) depending on the parameter $\alpha$ and $\theta$.}\label{Fig:ratio5}
\end{figure}

The success in reproducing the ratio $\Gamma[X\to  \chi_{cJ} \pi^0]/\Gamma[X\to  \pi^+ \pi^- J/\psi] $ with $J=1$ encourages us to apply the same mechanism to estimate the ratios for $J=0$ and $J=2$, which are presented in Figs.~\ref{Fig:ratio3} and \ref{Fig:ratio4}. Within the determined  $\theta$ range, the ratio $\Gamma[X\to  \chi_{cJ} \pi^0]/\Gamma[X\to  \pi^+ \pi^- J/\psi] $ are estimated to be $1.30 \sim 2.07$ and $1.12 \sim 1.28$ for $J=0$ and $J=2$, respectively, which indicate the  partial widths of $X(3872) \to \pi^0 \chi_{cJ}$ are very similar to each other. Furthermore, from Figs.~\ref{Fig:ratio2}-~\ref{Fig:ratio4}, one can find that the $\alpha$ and $\theta$ dependences of these ratios are also very similar, which are resulted form the similarity of $\chi_{cJ}, \ J=\{0,1,2\}$. In Fig.~\ref{Fig:ratio5}, we present the ratios $R_{01}=\Gamma[X\to \pi^0 \chi_{c0}]/\Gamma[X\to \pi^0 \chi_{c1}] $ and $R_{21}=\Gamma[X\to \pi^0 \chi_{c2}]/\Gamma[X\to \pi^0 \chi_{c1}] $ depending on the parameter $\alpha$ and $\theta$. In the determined $\theta$ range, the ratios are determined to be $\Gamma[X\to  \chi_{c0} \pi^0]:\Gamma[X\to  \chi_{c1} \pi^0]:\Gamma[X\to  \chi_{c2} \pi^0]=(1.77\sim 1.65):1:(1.09 \sim 1.43)$. This ratio is similar to the one estimated in an extended Friedrichs scheme, which is $1.5:1.3:1.0$~\cite{Zhou:2019swr}.

\section{Summary}
\label{sec:summary}
In the present work, we have investigated the decay behaviors of $X(3872)\to \pi^+ \pi^- J/\psi$, $\pi^+ \pi^- \pi^0 J/\psi$, and $\pi^0 \chi_{cJ}$ in a molecular scenario and tried to understand isospin violations in $X(3872)$ hidden charm decays. The fundamental source of the isospin violation has been shown in two different aspects. The mass difference of the charged and neutral charmed mesons leads to different coupling strengths of $X(3872) D^{\ast 0} \bar{D}^0$ and $X(3872) D^{\ast +} D^-$, which in part provides the source of isospin violation in the decay of $X(3872)$. Another important source of isospin violation is the interference between the charged %meson loops
and neutral meson loops.

By comparing our estimate of the ratio $\Gamma[X \to \pi^+ \pi^- \pi^0 J/\psi]/\Gamma[X \to \pi^+ \pi^- J/\psi] $ with the experimental data, we have determined the relative coupling strengths of $X(3872) D^{\ast 0} \bar{D}^0$ and $X(3872) D^{\ast +} D^-$. Then, with the fixed parameters, we have further estimated the ratios $\Gamma[X\to \pi^0 \chi_{cJ}]/\Gamma[X\to \pi^+ \pi^- J/\psi]$ with $J=0,1,2$. We have further tested our estimate by comparing our results of $J=1$ with the measurement from the BESIII collaboration and have found ours is well consistent with theirs. This fact has encouraged us to further predict the ratios for $J=0$ and $J=2$, which are $\Gamma[X\to \pi^0 \chi_{c0}]/\Gamma[X\to \pi^+ \pi^- J/\psi]=1.30 \sim 2.07$ and $\Gamma[X\to \pi^0 \chi_{c2}]/\Gamma[X\to \pi^+ \pi^- J/\psi]=1.12 \sim 1.28$, respectively. Moreover, in the determined parameter range, the partial width ratio of $\pi^0 \chi_{cJ}$ for $J=0,1,2$ is estimated to be $(1.77 \sim 1.65):1:(1.09 \sim 1.43)$, which could be tested by further precise measurements of BESIII and Belle II.

\section*{ACKNOWLEDGMENTS}
This project is partially supported by the National Natural Science Foundation of China under Grant No. 11775050. Qi Wu was supported by the Scientific Research Foundation of Graduate School of Southeast University (Grants no. YBPY2028).

\begin{appendix}
\section{Decay amplitudes}\label{sec:appendix}

The amplitudes for $X(3872)\to J/\psi \rho$ corresponding to diagrams in Fig.~\ref{Fig:Tri1} are
\begin{eqnarray}
\mathcal{M}_{a}&=&i^3 \int\frac{d^4 q}{(2\pi)^4}\Big[g_X \epsilon^X_\mu\Big]\Big[-ig_{\psi DD}\epsilon^{J/\psi}_\nu (-i)(p^\nu_1+q^\nu)\Big]\nonumber\\
&&\times\Big[-2f_{D^\ast DV}\varepsilon_{\rho\sigma\alpha\beta}ip^\rho_4 \epsilon^\sigma_{\rho/\omega} i(q^\alpha-p^\alpha_2)\Big]\nonumber\\
&&\frac{1}{p^2_1-m^2_1}\frac{1}{p^2_2-m^2_2}\frac{-g^{\mu\beta}+p^\mu_2 p^\beta_2 /m^2_2}{q^2-m^2_q}\mathcal{F}^2(q^2,m_q^2)\nonumber\\
%\end{eqnarray}
%\begin{eqnarray}
\mathcal{M}_{b}&=&i^3 \int\frac{d^4 q}{(2\pi)^4}\Big[g_X \epsilon^X_\mu\Big]\Big[g_{\psi D^\ast D}\varepsilon_{\rho\sigma\alpha\beta}ip^\rho_3 \epsilon^\sigma_{J/\psi}(-)(iq^\beta+ip^\beta_1)\Big]\nonumber\\
&&\times\Big[ig_{D^\ast D^\ast V}g^\nu_\tau g_{\theta\nu}(-i)(q_\kappa-p_{2\kappa})\epsilon^\kappa_{\rho/\omega}\nonumber\\
&&+4if_{D^\ast D^\ast V}g_{\tau\kappa}g_{\theta\nu}i(p^\kappa_4 \epsilon^\nu_{\rho/\omega}-p^\nu_4 \epsilon^\kappa_{\rho/\omega})\Big]\frac{1}{p^2_1-m^2_1}\nonumber\\
&&\frac{-g^{\mu\tau}+p^{\mu}_2 p^{\tau}_2 /m^2_2}{p^2_2-m^2_2}\frac{-g^{\alpha\theta}+q^{\alpha}q^{\theta}/m^2_q}{q^2-m^2_q}\mathcal{F}^2(q^2,m_q^2)\nonumber\\
%\end{eqnarray}
%\begin{eqnarray}
\mathcal{M}_{c}&=&i^3 \int\frac{d^4 q}{(2\pi)^4}\Big[g_X \epsilon^X_\mu\Big]\Big[g_{\psi D^\ast D}\varepsilon_{\rho\sigma\alpha\beta}ip^\rho_3 \epsilon^\sigma_{J/\psi}(-)(-ip^\beta_1-iq^\beta)\Big]\nonumber\\
&&\Big[-ig_{DDV}(-i)(p_{2\nu}-q_{\nu})\epsilon^\nu_{\rho/\omega}\Big]\frac{-g^{\mu\alpha}+p^{\mu}_1 p^{\alpha}_1 /m^2_1}{p^2_1-m^2_1}\nonumber\\
&&\frac{1}{p^2_2-m^2_2}\frac{1}{q^2-m^2_q}\mathcal{F}^2(q^2,m_q^2),\nonumber
\end{eqnarray}
\begin{eqnarray}
\mathcal{M}_{d}&=&i^3 \int\frac{d^4 q}{(2\pi)^4}\Big[g_X \epsilon^X_\mu\Big]\Big[ig_{\psi D^\ast D^\ast}\epsilon^\tau_{J/\psi}(-i)\Big(g_{\xi\theta}g_{\lambda\tau}(p^\theta_1+q^\theta)\nonumber\\
&&+g_{\xi\tau}g_{\lambda\theta}(p^\theta_1+q^\theta)-g_{\xi\theta}g^\theta_{\lambda}(p_{1\tau}+q_{\tau})\Big)\Big]\nonumber\\
&&\Big[2f_{D^\ast DV}\varepsilon_{\rho\sigma\alpha\beta}ip^\rho_4 \epsilon^\sigma_{\rho/\omega}(-i)(p^\alpha_2-q^\alpha)\Big]\frac{-g^{\mu\lambda}+p^{\mu}_1 p^{\lambda}_1 /m^2_1}{p^2_1-m^2_1}\nonumber\\
&&\frac{1}{p^2_2-m^2_2}\frac{-g^{\xi\beta}+q^{\xi} q^{\beta} /m^2_q}{q^2-m^2_q}\mathcal{F}^2(q^2,m_q^2),
\end{eqnarray}

The amplitudes for $X(3872)\to \chi_{cJ} \pi$ corresponding to diagrams in Fig.~\ref{Fig:Tri2} are
\begin{eqnarray}
\mathcal{M}_{a}&=&i^3 \int\frac{d^4 q}{(2\pi)^4}\Big[g_X \epsilon^X_\mu\Big]\Big[g_{\chi_{c0} DD}\Big]\Big[-ig_{D^\ast D P}ip_{4\nu}\Big]\nonumber\\
&&\frac{1}{p^2_1-m^2_1}\frac{-g^{\mu\nu}+p^\mu_2 p^\nu_2 /m^2_2}{p^2_2-m^2_2}\frac{1}{q^2-m^2_q}\mathcal{F}^2(q^2,m_q^2)\nonumber\\
%\end{eqnarray}
%\begin{eqnarray}
\mathcal{M}_{b}&=&i^3 \int\frac{d^4 q}{(2\pi)^4}\Big[g_X \epsilon^X_\mu\Big]\Big[g_{\chi_{c0} D^\ast D^\ast}\Big]\Big[-ig_{D^\ast D P}ip^\alpha_{4}\Big]\nonumber\\
&&\frac{-g^{\mu\nu}+p^{\mu}_1 p^{\nu}_1 /m^2_1}{p^2_1-m^2_1}\frac{1}{p^2_2-m^2_2}\frac{-g_{\nu\alpha}+q_{\nu}q_{\alpha}/m^2_q}{q^2-m^2_q}\mathcal{F}^2(q^2,m_q^2)\nonumber
\end{eqnarray}
\begin{eqnarray}
\mathcal{M}_{c}&=&i^3 \int\frac{d^4 q}{(2\pi)^4}\Big[g_X \epsilon^X_\mu\Big]\Big[-ig_{\chi_{c1} D^\ast D}\epsilon^{\chi_{c1}}_\nu\Big]\nonumber\\
&&\times\Big[\frac{1}{2}g_{D^\ast D^\ast P}\varepsilon_{\rho\sigma\alpha\beta}ip^\sigma_4 (-i)(p^\alpha_2-q^\alpha)\Big]\nonumber\\
&&\frac{1}{p^2_1-m^2_1}\frac{-g^{\mu\beta}+p^\mu_2 p^\beta_2 /m^2_2}{p^2_2-m^2_2}\frac{-g^{\nu\rho}+q^\nu
q^\rho /m^2_q}{q^2-m^2_q}\mathcal{F}^2(q^2,m_q^2),\nonumber\\
%\end{eqnarray}
%\begin{eqnarray}
\mathcal{M}_{d}&=&i^3 \int\frac{d^4 q}{(2\pi)^4}\Big[g_X \epsilon^X_\mu\Big]\Big[g_{\chi_{c2} D^\ast D^\ast}\epsilon^{\chi_{c2}}_{\alpha\beta}\Big]\Big[-ig_{D^\ast D P}ip_{4\nu}\Big]\nonumber\\
&&\frac{-g^{\mu\alpha}+p^{\mu}_1 p^{\alpha}_1 /m^2_1}{p^2_1-m^2_1}\frac{1}{p^2_2-m^2_2}\frac{-g^{\beta\nu}+q^{\beta} q^{\nu} /m^2_q}{q^2-m^2_q}\mathcal{F}^2(q^2,m_q^2),\nonumber\\
\end{eqnarray}

\end{appendix}

\end{document}